\documentstyle[prb,aps,epsf,multicol]{revtex}

\begin{document}
\draft 

\title{Glassy Vortex State in a Two-Dimensional Disordered XY-Model} 

\author{Stefan Scheidl} 

\address{Universit\"at zu K\"oln, Institut f\"ur Theoretische Physik,
  Z\"ulpicher Str. 77, D-50937 K\"oln, Germany}

\date{\today}
\maketitle

\begin{abstract}
  The two-dimensional XY-model with random phase-shifts on bonds is
  studied. The analysis is based on a renormalization group for the
  replicated system. The model is shown to have an ordered phase with
  quasi long-range order. This ordered phase consists of a glass-like
  region at lower temperatures and of a non-glassy region at higher
  temperatures. The transition from the disordered phase into the
  ordered phase is not reentrant and is of a new universality class at
  zero temperature. In contrast to previous approaches the disorder
  strength is found to be renormalized to larger values.  Several
  correlation functions are calculated for the ordered phase.  They
  allow to identify not only the transition into the glassy phase but
  also an additional crossover line, where the disconnected vortex
  correlation changes its behavior on large scales non-analytically.
  The renormalization group approach yields the glassy features
  without a breaking of replica symmetry.
\end{abstract}


\begin{multicols}{2}\narrowtext

\section{Introduction}

The two-dimensional XY-model with random phase-shift, defined in Eq.
(\ref{H}) below, captures the physics of a variety of physical
systems. Among them are XY-magnets with non-magnetic impurities, which
are coupled to the XY-spins via the Dzyaloshinskii-Moriya
interaction,\cite{RSN83} Josephson-junction arrays with geometric
disorder,\cite{GK} and crystals with quenched impurities.\cite{drN83}
Disordered XY-models are related to even more physical systems, like
impure superconductors. It is in particular the interest in
vortex-glass phases in type-II superconductors, which motivates an
analysis of possible glassy features in the paradigmatic
two-dimensional bond-disordered system.

The most fundamental question is, whether this model has an ordered
phase. In the absence of disorder, there exists at low temperatures a
phase with quasi long-range order. The famous Kosterlitz-Thouless
transition separates it from the high-temperature phase with
short-ranged order.\cite{KT73,jmK74} The analysis by Cardy and
Ostlund\cite{CO82} and more explicitly by Rubinstein, Shraiman and
Nelson\cite{RSN83} predicted that the phase transition should be
reentrant in the presence of disorder. Later Korshunov
suggested,\cite{seK93} that the ordered phase might be completely
destroyed by disorder. Ozeki and Nishimori\cite{ON93} showed for
models with gauge-invariant disorder distributions, that the phase
transition cannot be reentrant, leaving open whether the ordered phase
does exist or not.  Experiments\cite{exp} on Josephson junction arrays
and simulations\cite{sim} were in favor of the existence of an ordered
phase without reentrance.

In this controversial situation Nattermann, Scheidl, Korshunov, and
Li\cite{NSKL95} as well as Cha and Fertig\cite{CF95} reconsidered only
recently the problem and found that the ordered phase {\em does} exist
and its boundary {\em does not} have reentrant shape. The earlier
observation of reentrance was attributed to an overestimation of
vortex fluctuations at low temperatures.

The present study extends Refs. \CITE{NSKL95} and \CITE{CF95},
which focused on the absence of reentrance. The purpose of this
article is to examine in more detail the ordered phase, which is shown
to be composed of different {\em regions}. They are distinguished by
the behavior of vortex correlation functions on large length scales,
which are calculated {\em quantitatively}. A low-temperature regime is
found to exhibit glassy behavior in its correlations.

In addition to the previous studies we find that the strength of
disorder is increased under renormalization. This effect does not
destroy the ordered phase but it is crucial for the universality type
of the transition at low temperatures.

In the following a generalized self-consistent screening approach is
applied to the replicated system. The main improvement compared to
earlier approaches\cite{RSN83,CO82,seK93} on the same basis is
achieved by including all relevant fluctuations and by taking special
care when the number of replicas is sent to zero.

We use the replica formalism mainly for two reasons: i) The
calculation of correlations becomes very convenient: one may use an
expansion technique (small fugacity and density of vortices) which
fails\cite{NSKL95,CF95,KN95} in the unreplicated system. ii) We show
that in the present model there is {\em no} need for replica symmetry
breaking, even at lowest temperatures. In particular, the reentrance
disappears and the glassiness occurs without a breaking of replica
symmetry. This might astonish, since for various disordered systems
reentrance has been explained as an artifact of a replica-symmetric
(exact mean-field or approximate variational) approach, which has been
overcome by breaking the replica symmetry.\cite{RSB+reen}

The paper is organized as follows: In Sec. II we set up our model and
map it onto an effective Coulomb gas. We recapitulate the breakdown of
fugacity expansion in the presence of disorder. In Sec. III we set up
the renormalization group within a self-consistent formulation of
screening. The consistency of this approach in the original and
replicated system is demonstrated. Sec. IV is devoted to the
evaluation of the self-consistency and the proper performance of the
replica limit. The results of the renormalization group treatment are
worked out in Sec. V, which are discussed and compared to the results
of previous work in Sec. VI.

\section{The Model}

The XY-model with random phase-shifts is given by the reduced
Hamiltonian
\begin{equation} \label{H} 
  {\cal H}=-K \sum_{<{\bf r, r}'>} \cos (\theta_{\bf r}-\theta_{\bf
    r'} -A_{<{\bf r, r}'>}) .
\end{equation} 
It refers to a square lattice with unit spacing in two dimensions. An
XY-spin with angle $\theta_{\bf r} \in [-\pi,\pi[$ is attached to
every lattice site (position ${\bf r}$). The reduced spin coupling
between nearest neighbors $<{\bf r, r}'>$ reads $K=J/T$. The
interaction involves quenched random phase-shifts with variance
$\overline{A_{<{\bf r, r}'>}^2}=\sigma$, which are uncorrelated on
different bonds. Thermal fluctuations are weighted according to the
partition sum
\begin{equation} \label{Z}
  Z=\int_{-\pi}^\pi {\cal D}\{\theta \} \ e^{-{\cal H}} .
\end{equation}

This XY-model can be mapped approximately onto an effective Coulomb
gas. Since this procedure is well known in the absence of
disorder\cite{Vil75,JKKN77} and can be performed similarly in the
presence of disorder, we recall only briefly the main manipulations
leading to the effective model. In order to simplify the analysis of
model (\ref{Z}), one can replace approximately the cosine-interaction
by the Villain-interaction.\cite{Vil75} Thereby an additional
variable, the vortex density $N$ is introduced. The decisive advantage
of this approximation is, that the Hamiltonian becomes bilinear in the
angles. All configurations of the original angles can be expressed in
terms of spin-waves and vortices. Just as in the pure case, spin-waves
and vortices decouple energetically. Since spin-waves are simply
harmonic, all non-trivial physics arises from vortices with an
effective Hamiltonian
\begin{mathletters}
\label{H.eff}
\begin{eqnarray}\label{H.eff.a} 
  {\cal H}_{\rm v}&=& \int_{\bf k} \frac {2 \pi^2 K } {k^2}
  \left|N_{\bf k}-Q_{\bf k} \right|^2
\label{Hv.k}\\
&=& E\sum_{\bf R} (N_{\bf R}-Q_{\bf R})^2 - \label{Hv.R} \\ &&-
\sum_{\bf R \neq R'} \pi K \ln |{\bf R}- {\bf R}'| (N_{\bf R}-Q_{\bf
  R})(N_{\bf R'}-Q_{\bf R'}) ,\nonumber
\end{eqnarray}
\end{mathletters}
in Fourier and real space, and the partition sum
\begin{equation} \label{Z.vc}
  Z_{\rm v}= \sum_{\{N\}} e^{-{\cal H}_{\rm v}} .
\end{equation}
Here $N_{\bf R}$ is the integer-valued vortex density, which is
subject to thermal fluctuations. On the square lattice $N_{\bf R}$ is
associated with plaquettes (dual sites), which are labeled by ${\bf
  R}$ in contrast to sites ${\bf r}$. {\em Thermal} vortices $N$
couple to a background of {\em quenched} vortices $Q_{\bf R}=\frac 1{2
  \pi} {\nabla} \times {\bf A}$. The last expression means, that the
plaquette variable $Q_{\bf R}$ is given by the rotation of variable
$A$ on the surrounding bonds. From this relation one immediately
derives, that the quenched vortices are Gaussian random variables with
variance $\overline{Q_{-\bf k} Q_{\bf k}}= \frac {\sigma}{4 \pi^2}
k^2$. A single phase-shift $A_{<{\bf r, r}'>}$ creates a dipole of
quenched vortices situated in the plaquettes which have the bond $
<{\bf r},{\bf r}'>$ as side. Therefore quenched vortices are
anti-correlated and their total vorticity vanishes.

We recognize from Eq. (\ref{Hv.k}), that only neutral configurations
of vortices will have finite energy. Hence the partition sum can be
restricted to neutral configurations (i.e.  $\sum_{\bf R} {\bf N}_{\bf
  R} = 0$) like in the pure case. Every such neutral configuration
can be imagined as a superposition of dipoles of a vortex ($N=1$) and
an antivortex ($N=-1$). In analogy to electrostatics, vortices of the
same sign repel each other and vortices of different sign attract each
other.

Since the interaction is logarithmic, vortices build a Coulomb gas.
The interaction is identical among the thermal and quenched component.
The Fourier representation (\ref{Hv.k}) of the vortex Hamiltonian
makes evident, that thermal vortices try to compensate quenched
vortices as well as possible. A perfect compensation is prevented by
the fact, that $N_{\bf R}$ is restricted to integer values, whereas
$Q_{\bf R}$ is a continuous random variable.

In Eq. (\ref{Hv.R}) we introduced the vortex core energy $E=\pi \gamma
K$, where $\gamma$ takes the value $\gamma \approx 1.6$ on a square
lattice. It emerges from the lattice Greens function [cf.  Eq.
(4.13a) of Ref. \CITE{JKKN77}]. Although $\gamma$ has a unique value
for the lattice model, it will occasionally be considered as an
independent parameter which allows to control the density of vortices
irrespectively of $K$. In a dilute vortex system with $N_{\bf R}= 0,
\pm 1$ only, $E$ is comparable to a chemical potential.

In a region of parameters, where vortices are negligible, fluctuations
of the original angle variables are given only by spin waves. They
lead to a decay of the spin correlation function\cite{RSN83,CO82}
\begin{equation} \label{def.Gamma}
  \Gamma({\bf r}) = \overline{\langle \cos[ \theta({\bf r}) -
    \theta({\bf 0})] \rangle} \sim r^{- \eta}
\end{equation}
with exponent $\eta={(1/K+\sigma) /2\pi }$. If the correlation
decays algebraically even in the presence of vortices, the system has
quasi long-range order.

As vortices are excited thermally, they give rise to additional
fluctuations in the angles, the correlation will decay faster. This
effect can be expressed by a renormalized $\eta$, which becomes scale
dependent. If vortices lead to a divergence of the renormalized $\eta$
on large scales, quasi long-range order is destroyed. However, vortex
fluctuations are not only responsible for such a quantitative effect,
they also drive the phase transition into the high-temperature phase,
where the spin-spin correlation decays exponentially.\cite{KT73}

Let us estimate, to what extent disorder can favor the excitation of
vortices. In the pure case $Q=0$, the ground state of (\ref{Hv.k}) is
vortex-free. The density of vortices vanishes for vanishing
temperature, since the creation of quantized vortices always costs
finite energy. Therefore the pure system has a phase with quasi
long-range order (\ref{def.Gamma}) at low temperatures.

A simple argument shows, that the {\em disordered} system has even at
{\em zero} temperature a {\em finite density} of
vortices:\cite{NSKL95,CF95} we consider two positions at distance $R$
and determine the probability of finding there a vortex-antivortex
pair. The energy of such a dipole is given by $U_{\rm dipole}(R)
\approx 2 \pi J \ln R$, where we neglect the core energy for large
$R$. Disorder gives and additional energy contribution $V_{\rm
  dipole}(R)$. At $T=0$, the dipole will be present, if its total
energy $U_{\rm dipole}(R)+V_{\rm dipole}(R)$ is negative. The original
Gaussian distribution of the random phase-shifts results in a Gaussian
distribution of $V_{\rm dipole}(R)$ with variance $\Delta^2(R) :=
\overline {V_{\rm dipole}^2(R)} \approx 4\pi \sigma J^2 \ln R $. Thus
the probability
\begin{eqnarray}\label{prob.dipole}
  P(R)&=&\int_{-\infty}^{-U_{\rm dipole}(R)} \frac{dV}{\sqrt{2 \pi
      \Delta(R)}} \ e^{-V^2/2 \Delta^2(R)} \nonumber \\ &\approx&
  \sqrt{\frac{\sigma}{2 \pi^2 \ln R}} \ R^{-\pi/2 \sigma}
\end{eqnarray}
of finding a pair at $T=0$ is finite. In other words, this shows that
in the presence of disorder the ground state is no longer vortex-free
and that renormalization effects can be important even at arbitrarily
small temperatures.

\section{Screening}

The goal of the present work is therefore a calculation of such
renormalization effects in the presence of random phase-shifts. We use
the conceptual most simple approach, a self-consistent linear response
approach. For the pure Coulomb gas, this approach is
equivalent\cite{apY78} to the real-space renormalization group of
Kosterlitz.\cite{jmK74} We have checked that this equivalence still
holds in the presence of disorder.

We now define vortex renormalization effects by macroscopic properties
of our system. As a probe we introduce additional test vortices into
the system. They experience a screened interaction and the screened
background potential acting on them. Let us denote the unscreened
interaction between vortices $U_{\bf k}= 4 \pi^2 K \frac{1}{k^2}$. Due
to quenched disorder vortices $Q$, thermal vortices $N$ are subject to
a Gaussian background potential $V_{\bf k}= U_{\bf k} Q_{\bf k}$ with
variance $\overline{V_{-\bf k} V_{\bf k}}= 4 \pi^2 \sigma K^2 \frac
1{k^2}$. The screened background potential and interaction are
identified from the contributions to the free energy of test vortices,
which are of first and second order in their vorticity\cite{screen}
(see also Appendix \ref{app.lr} for some intermediate steps)
\begin{mathletters}\label{screen.V}
\begin{eqnarray}\label{screen.V.a}
  V_{\bf k}^{\rm scr} &=& U_{\bf k} (Q_{\bf k} - \langle N_{\bf k}
  \rangle), \\ U_{\bf k}^{\rm scr} &=& U_{\bf k} - U_{\bf k}^2 C_{\rm
    con}({\bf k}).
\end{eqnarray}
\end{mathletters}
It is convenient to introduce the ordinary, disconnected, and
connected correlation,
\begin{mathletters}\label{def.corr}
\begin{eqnarray}\label{def.corr.a}
  C({\bf R}) &:=&\overline{\langle N_{\bf R} N_{\bf 0} \rangle},\\ 
  C_{\rm dis}({\bf R}) &:=& \overline{\langle N_{\bf R}\rangle \langle
    N_{\bf 0} \rangle},
\label{def.C.neq}\\
C_{\rm con}({\bf R})&:=& C({\bf R})-C_{\rm dis}({\bf R}).
\end{eqnarray}
\end{mathletters}
For a specific realization of disorder, the screened interaction
in principle depends explicitly on the coordinates of both interacting
vortices, it is not translation invariant. Although we should in
principle consider screening in the particular disorder environment,
we approximately evaluate only the average screening effect by taking
the disorder average, which restores the translation symmetry of the
interaction.

From the large-scale behavior of the screened interaction and the
variance of the screened potential one can then identify screened
parameters $K^{\rm scr}$ and $\sigma^{\rm scr}$ as (for the precise
definition see Appendix \ref{app.lr})
\begin{mathletters}\label{total.screen}
\begin{eqnarray}\label{total.screen.a}
  K^{\rm scr} &=& K + 2 \pi^3 K^2 \int_1^{\infty} dR R^3 C_{\rm
    con}(R),
\label{total.screen.K} \\
\sigma^{\rm scr}(K^{\rm scr})^2 &=& \sigma K^2 - 2 \pi^3 K^2
\int_1^{\infty} dR R^3 C_{\rm dis}(R) + \nonumber \\ && + 4 \pi^4
\sigma K^3 \int_1^{\infty} dR R^3 C_{\rm con}(R).
\label{total.screen.sig}
\end{eqnarray}
\end{mathletters}

Although we do not yet know the correlation functions, we may
establish from these expressions criteria for the existence of an
ordered phase. In this phase, we expect vortices to give rise only to
finite screening effects, i.e. the screened parameters must have
finite values. This requires $\lim_{R \to \infty} R^4 C_{\rm con}(R) <
\infty$ and a similar condition for the disconnected correlation.

In the limit of large core energy, one can calculate the correlations
to leading order by considering only a single vortex dipole. In th
absence of disorder, one finds\cite{} $C_{\rm con}({\bf R})= C({\bf
  R}) \sim R^{-2 \pi K}$ and $C_{\rm dis}({\bf R})=0$. The condition
for the ordered phase then simply reads $K \geq 2/\pi$, in agreement
with the condition that the free energy a single vortex has to be
positive.\cite{KT73} A similar argument can be constructed for the
disordered system at $T=0$, where one can estimate $C(R) \sim p(R)\sim
R^{-\pi/2 \sigma}$ using Eq. (\ref{prob.dipole}). Then the condition
for order reads $\sigma \leq \pi/8$. This argument already disproves
previous predictions,\cite{RSN83,CO82,seK93} that infinitesimal
$\sigma > 0$ would destroy order at $T=0$.

For finite core energy, screening effects have to be taken into
account quantitatively for the calculation of the correlations. For
this purpose we develop a self-consistent scheme in analogy to the pure
case.\cite{KT73,screen} We introduce scale-dependent variables
$K\equiv K(l)$ and $\sigma \equiv \sigma(l)$, which include screening
by dipoles of radius $1 \leq R < e^l$ only. Then Eq.
(\ref{total.screen}) can be cast in differential form:
\begin{mathletters}\label{partial.screen}
\begin{eqnarray}\label{partial.screen.a}
  \frac{d}{dl} K^{-1} &=& -2 \pi^3 e^{4l} C_{\rm con}(R=e^l),
\label {partial.screen.K}\\
\frac{d}{dl} \sigma &=& -2 \pi^3 e^{4l} C_{\rm dis}(R=e^l).
\label {partial.screen.sig}
\end{eqnarray}
\end{mathletters}
Both equations combine to a simple flow equation of the spin-spin
correlation exponent, $ \frac{d}{dl} \eta = - \pi^2 e^{4l} C(R=e^l)$.

Eqs. (\ref{partial.screen}) suggests a simple ``two-component
picture'' illustrating the screening effects of vortices. This picture
virtually separates dipoles of thermal vortices $N$ into a frozen and
a polarizable component. The {\em frozen} component gives rise to the
$C_{\rm dis}$.  It is not polarizable and does not contribute to
screening of the interaction. However, since it is frozen, it behaves
like the quenched disorder background and leads to an effective
disorder strength represented by the screened $\sigma$. The {\em
  polarizable} component is the source of $C_{\rm con}$ and leads to
the screening of $K$. Both components contribute equally to a
renormalization of the spin exponent. This virtual separation into two
components must not be taken literally. It is impossible to assign a
particular dipole uniquely to one of the two components. The
two-component picture is analogous to the two-fluid model of
superfluidity, which also must not be taken literally in the sense
that a particular atom would be either superfluid or normalfluid.

In order to take advantage from the differential screening equations,
we will calculate the correlations at $R=e^l$ in a self-consistent way
which includes screening effects by smaller dipoles. At each
differential step we will also perform a simultaneous rescaling $R \to
e^{-dl} R $. Then we can interpret the differential equations as
renormalization group flow equations.

For the further investigation of the model, we apply the standard
replica technique\cite{EA75} to the partition function (\ref{Z.vc}).
We are going to rewrite screening in the replicated system in order to
expose the consistency between screening in the unreplicated and
replicated system.

The Coulomb gas (\ref{Hv.R}) is replicated $n$ times and after
disorder averaging one obtains the effective Hamiltonian
\begin{eqnarray}\label{H.rep} 
  {\cal H}_n &=& \sum_{\bf R} \sum_{ab} E^{ab} N_{\bf R}^a N_{\bf R}^b
  - \nonumber \\ &&-\frac 12 \sum_{{\bf R} \neq {\bf R}'} \sum_{a,b} 2
  \pi K^{ab} \ln|{\bf R}-{\bf R}'| N_{\bf R}^a N_{\bf R'}^b \ .
\end{eqnarray}
We introduce the couplings $K^{ab}=K \delta^{ab}- \hat{K}$,
$\hat{K}=\sigma K^2/ ({1+n \sigma K})$ and the core energy $E^{ab}=\pi
\gamma K^{ab}=E \delta^{ab}- \hat{E}$, where $E=\pi \gamma K$ and
$\hat{E}=\pi \gamma \hat{K}$. On the partition sum the neutrality
condition
\begin{equation} \label{neutrality}
  \sum_{\bf R} {\bf N}_{\bf R}^a = 0
\end{equation}
is imposed for every replica.

In the replicated system, screening effects can be calculated in the
same linear response scheme as before. The leading term of the free
energy of test replica-vortices is of second order in the
infinitesimal test vorticity. From this order, we derive a
differential screening
\begin{equation} \label{partial.screen.rep}
  \frac{d}{dl} K^{ab} = 2 \pi^3 e^{4l} \sum_{cd} K^{ac} C^{cd}(R=e^l)
  K^{db}
\end{equation}
in analogy to Eq. (\ref{partial.screen}). We introduced the replica
correlation $C^{ab}(R):=\langle N^a_{\bf R} N^b_{\bf 0}\rangle_n$,
where $\langle \dots \rangle_n$ denotes a thermal average in the
replicated system. Since the initial interactions are
replica-symmetric, i.e. the coupling has the form $K^{ab}=K
\delta^{ab} - \hat{K}$, we use the replica symmetric ansatz
$C^{ab}=C_{\rm con}^{(n)} \delta^{ab} + C^{(n)}_{\rm dis}$ for
correlations. The relation to the correlations in the unreplicated
system is provided by
\begin{equation} \label{rel.corr}
  C_X({\bf R}) = \lim_{n \to 0} C^{(n)}_X({\bf R})
\end{equation}
for all three types $X$ (ordinary, disconnected, and connected) of
correlations, if we identify $C^{(n)}({\bf R}):= C^{aa}({\bf R})$
consistently.

Exploiting the replica symmetry of correlations, Eq.
(\ref{partial.screen.rep}) decays into flow equations for $K$ and of
$\hat{K}$, which read in terms of $K$ and $\sigma=\hat{K} / (K^2- n K
\hat{K})$:
\begin{mathletters}\label{partial.screen.ren}
\begin{eqnarray}\label{partial.screen.ren.a}
  \frac{d}{dl} K^{-1} &=& 4 \pi^3 y_{\rm con}^2,
\label {partial.screen.K.ren}\\
\frac{d}{dl} \sigma &=& 4 \pi^3 y_{\rm dis}^2.
\label {partial.screen.sig.ren}
\end{eqnarray}
\end{mathletters}

In order to approach the usual notation of renormalization groups, we
define {\em effective} fugacities
\begin{equation} \label{def.y.eff}
  y^2_X \equiv y^2_X(l) := - \frac 12 e^{4l} C^{(n)}_X(R=1)
\end{equation}
again for all three types of rescaled correlations. They are related
by $y^2=y_{\rm dis}^2+y_{\rm con}^2$. In these definitions we took
already account of rescaling of lengths, which generates an additional
flow of the core energies,
\begin{mathletters}\label{flow.E}
\begin{eqnarray}\label{flow.E.a}
  \frac{d}{dl} E &=& \pi K , \\ \frac{d}{dl} \hat{E} &=& \pi \hat{K} .
\end{eqnarray}
\end{mathletters}

Eqs. (\ref{partial.screen.ren}) together with (\ref{flow.E})
constitute the main renormalization group flow equations. From
relations (\ref{rel.corr}) we recognize, that the flow equations of
$K$ and $\sigma$ of the unreplicated system coincide with those of the
replicated system in the limit $n \to 0$. In the next section these
equations will be completed by a flow equation of the free energy,
which however does not feed into the other flow equations.

\section{Self-consistent closure}

The flow equations (\ref{partial.screen.ren}) together with
(\ref{flow.E}) can not yet be evaluated since they are not closed. In
this section we are going to achieve this closure by expressing the
effective fugacities in terms of $K$, $\sigma$, $E$ and $\hat{E}$.

In the replica language, the effect of disorder is encoded in
additional interactions between vortices and we first have to discuss
their nature. An inspection of Eq. (\ref{H.rep}) shows, that vortices
of opposite sign in the same replica interact with a potential $2 \pi
(K-\hat{K}) \ln R$. For $n \geq 1$ they always attract each other.
These dipoles are stable against dissociation due to thermal
fluctuations only for $(K-\hat{K}) \geq 2/\pi$.\cite{KT73,jmK74} In
previous work \cite{RSN83,CO82} the stability of precisely such
dipoles was used as criterion to determine the boundary of the ordered
phase. However, for $n=0$ and low temperature or strong disorder,
$K-\hat{K}=K-\sigma K^2$ becomes negative (repulsion!) and this
criterion becomes questionable. The increasing instability of these
dipoles gives rise to a reentrant phase boundary.\cite{RSN83}

However, it is not sufficient to consider interactions only within
replicas. Independently on temperature and replica number $n$,
vortices of the same sign but in different replicas attract each other
with a potential $2 \pi \hat{K} \ln R$. In the limit $n=0$ the
coupling between different replicas reads $\hat{K}=\sigma K^2$. For
low temperature or strong disorder, the interaction between different
replicas becomes stronger to the same extent as the intra-replica
coupling becomes weaker and correlations have to be calculated taking
into account the competition between intra- and inter-replica
interactions. The importance of inter-replica interactions has been
recognized by Korshunov.\cite{seK93} 

The essence of the replica problem is a suitable treatment of the
inter-replica interaction for integer $n \geq 1$ and to construct an
analytic continuation which allows to take then the limit $n\to 0$ in
a proper way. In the following this is done in a way different from
Korshunov's way, leading to opposite conclusions. On the technical
level, this step is the essential progress of the present work.

\subsection{Physics of $n\geq 1$}

As necessary prerequisite for the analytic continuation $n \to 0$, we
must discuss vortex fluctuations in the replicated system for integer
$n \geq 1$. This happens in some detail since our final conclusions
contradict Refs.~\CITE{RSN83,CO82,seK93}.

We first check, that the {\em ground state} of the {\em replicated}
Hamiltonian is the state without vortices. For this purpose we examine
the Hamiltonian in Fourier space, ${\cal H}_n= \int_{\bf k} \sum_{ab}
(2 \pi^2 /k^2) K^{ab} N^a_{-\bf k} N^b_{\bf k}$. There we easily
recognize, that every creation of vortices costs energy, because
$K^{ab}$ is positive definite: $K^{ab}$ has one eigenvalue
$\kappa_1=K-\hat{K}=K(1+(n-1)\sigma K)/(1+n\sigma K)>0$ and $n-1$
eigenvalues $\kappa_{2,\dots,n}=K>0$. Since the ground state has no
vortices, we are allowed to use for low temperatures a low density
expansion. This is in contrast to the unreplicated system, where the
ground state has a finite density of vortices, as discussed above.

\begin{figure}
\epsfxsize=\linewidth
\epsfbox{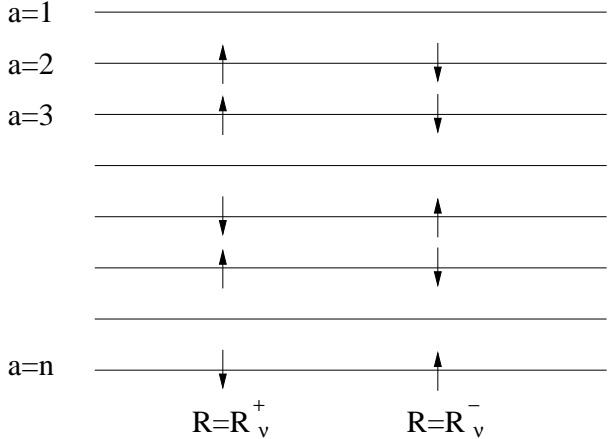}
\caption{
  Schematic picture of a dipole of replica-vortices. Every horizontal
  line represents a two-dimensional system, which is replicated $n=8$
  times in this example. An up/down arrow represents a vorticity
  $N_{\bf R}=\pm 1$. The dipole consists of a vortex of type
  $\nu=\{0,1,1,0,-1,1,0,-1\} $ and class $(m_0=5,m_1=3)$ and its
  antivortex at positions ${\bf R}={\bf R}_\nu^\pm$.}
\label{fig_dipole}
\end{figure}

For the following it is convenient to introduce the notion of a
``replica-vortex'' (compare Fig. \ref{fig_dipole}): in a state with
vorticity $\{ N^{1}_{\bf R},\dots,N^{n}_{\bf R} \}$ at position ${\bf
  R}$, we say that a ``replica-vortex'' of {\em type} $\nu \equiv \{
N^{1}_{\bf R},\dots,N^{n}_{\bf R} \}$ is at position ${\bf R}$.
(Upper indices at $N$ are not exponents but replica indices!) Since
the core energy contains a term $E \sum_{{\bf R}a} (N^a_{\bf R})^2$,
which suppresses large $|N^a|$, we may restrict our consideration for
large $E$ or $\gamma$ to states with $N^a_{\bf R}=0,\pm 1$: one
vortex-free state and $3^n-1$ different types of replica-vortices,
antivortices included.

Because of the neutrality condition (\ref{neutrality}) the elementary
excitations in the replicated system are again vortex dipoles. In the
dilute limit (large $\gamma$), some type of such replica-vortices will
form with their antivortices those dipoles, which are most instable
against thermal fluctuation and thus destroy order. To find out this
type, we have to discuss the energy of such dipoles. The core energy
$E_{\nu}$ of replica-vortex $\nu = \{ N^{1},\dots,N^{n}\}$ depends
only on the total vortex number $m_0:=\sum_a (N^a)^2$ and the number
$m_1:=\frac 12 \left(m_0+\sum_a N^a \right)$ of vortices with
$N^a=+1$.  The same holds for the coupling $K_{\nu}$, which gives the
strength of the logarithmic interaction between a replica-vortex of
type $\nu$ and its antivortex:
\begin{mathletters}\label{def.E_m}
\begin{eqnarray}\label{def.E_m.a}
  E_{\nu}=m_0 E-(m_0-2m_1)^2 \hat{E} ,\\ K_{\nu}=m_0 K-(m_0-2m_1)^2
  \hat{K} .
\end{eqnarray}
\end{mathletters}
Due to replica symmetry, there is a certain degeneracy between
different types $\nu$, which we include when we speak of the {\em
  class} $(m_0,m_1)$ of replica vortices.

\begin{figure}
\epsfxsize=\linewidth
\epsfbox{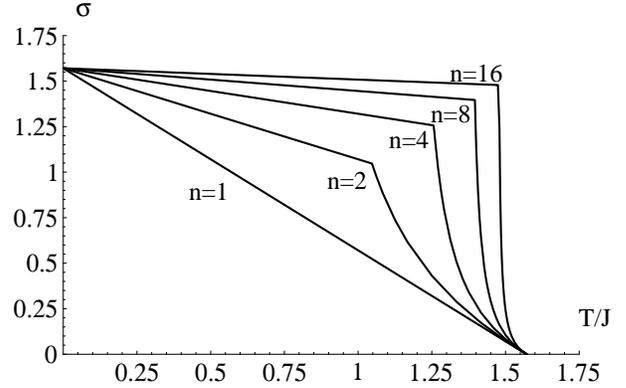}
\caption{
  Phase diagram for $n=1,2,4,8,16$ in the limit $\gamma \to \infty$.
  Below the transition line all types of dipoles are stable against
  thermal dissociation, above the line there is at least one instable
  type.  At high temperatures (above the shoulder of the transition
  line), the transition is driven by the dissociation of replica
  dipoles $(m_0=1,m_1=0,1)$, whereas it is driven by dipoles
  $(m_0=n,m_1=0,n)$ at low temperatures (below the shoulder of the
  transition line).}
\label{fig_pdgn}
\end{figure}

The existence of order requires stability of {\em all} types of
dipoles against thermal dissociation. For the dilute system this is
guaranteed by $\min_{\nu\neq 0} K_\nu \geq 2\pi$ according to the
simple stability criterion of Kosterlitz and Thouless.\cite{KT73} The
least stable class with the weakest coupling can be determined
explicitly: it is given by replica-vortices of class $(m_0=1,m_1=0,1)$
with $K_\nu=K-\hat{K}$ in a high temperature region $(n+1) \hat{K}
\leq K$, and by replica-vortices of classes $(m_0=n,m_1=0,n)$ with
$K_\nu=n K-n^2\hat{K}$ in a low temperature region $(n+1) \hat{K} \geq
K$. The resulting phase diagram for $n=1,2,4,8,16$ is shown in Fig.
\ref{fig_pdgn}.

The fact, that the class of the least stable replica-vortices for $n >
1$ depends on the physical parameters, can be interpreted as an
indication for a phase transition even for $n=0$. We take it as a
warning, that we must not focus on a {\em single} class of vortices.
The origin for the failure of the early replica
approaches\cite{RSN83,CO82} is just the focus on classes
$(m_0=1,m_1=0,1)$ only. We even do not restrict ourselves to the two
classes which are dominant for $n>1$, since it turns out {\em a
  posteriori}, that also other classes contribute to the replica limit
$n \to 0$ .

\subsection{Towards the replica limit $n \to 0$}

Now we calculate the correlations and the effective fugacities taking
into account contributions by all types of replica-vortices. For a low
density of replica-dipoles (realized for large $\gamma$ or at low
temperature) we can neglect the interaction between different
replica-dipoles (``independent dipole approximation''). This means
that we can calculate approximate correlations in the partition sum
containing only up to one dipole of replica-vortices,
\begin{equation} \label{Z.lo}
  Z^{(n)}=1+\frac 12 \sum_{\nu\neq 0} y_\nu^2 \int d^2 R_{\nu}^+ d^2
  R_{\nu}^- |{\bf R}_{\nu}^+ -{\bf R}_{\nu}^-|^{-2 \pi K_\nu} .
\end{equation}
The fugacity of replica-vortex $\nu$ is defined by $y_\nu :=
e^{2l-E_{\nu}}$. The factor $1/2$ is present, since $\nu$ runs over
vortices and their antivortices and we should count every realization
of a dipole only once. In Eq. (\ref{Z.lo}) integration is restricted
to $|{\bf R}_{\nu}^+-{\bf R}_{\nu}^-| \geq 1$ for all $l \geq 0$,
since lengths are rescaled.

Before we turn to the calculation of correlations, we wish to make
sure that calculations in this ensemble make sense in the limit $n \to
0$. For this purpose we have to convince ourselves, that $Z_n \to 1$.
This is equivalent to the condition, that the free energy per replica
has a finite limit. Then also the renormalization group flow of the
free energy per replica, which follows from the contribution of
dipoles of radius $1\leq R < e^{dl}$ to $Z_n$, is finite. The flow
reads
\begin{equation} \label{flow.F}
  \frac{d}{dl} {\cal F}= 2 {\cal F} - \pi \frac 1n \sum_{\nu \neq 0}
  y_\nu^2 \ ,
\end{equation}
where ${\cal F}$ denotes the free energy per unit volume, per replica,
and divided by temperature. The first term on the right-hand side
originates from rescaling.

From Eq. (\ref{flow.F}) one can derive (see Appendix
\ref{app.extreme}), that a restriction to a single class of
replica-vortices leads to an inconsistency with the replica trick,
namely the divergence of ${\cal F}$ for $n \to 0$. Therefore we retain
{\em all} classes of vortices.

Before we turn to the correlations, we evaluate the simpler sum, which
is the source for the flow of free energy:
\begin{eqnarray}\label{sum.y}
  \sum_{\nu \neq 0} y_\nu^2 &=& e^{4l} \sum_{\{N^a\}\neq 0} e^{ - 2 E
    \sum_a \left( N^a \right)^2 + 2 \hat{E} \left(\sum_a N^a\right)^2
    } \nonumber \\ &=& e^{4l} \sum_{\{N^a\}\neq 0} \overline{ e^{ - 2
      E \sum_a \left( N^a \right)^2 + 2 {\cal A} \sqrt{2 \hat{E}}
      \sum_a N^a } } \nonumber \\ &=& e^{4l} \left\{\overline{z^n}-1
  \right\} ,
\end{eqnarray}
where we introduced a ``dummy'' Gaussian random variable ${\cal A}$
with $\overline{{\cal A}}=0$ and $\overline{{\cal A}^2}=1/2$ and a
``shell partition sum'' $z \equiv z({\cal A},E,\hat{E}):=1+z_+ +z_-$
with weights
\begin{equation} \label{def.zpm}
  z_\pm : = e^{-2(E \pm {\cal A} \sqrt{2 \hat{E}})}.
\end{equation}

Physically, the ``dummy'' Gaussian random variable ${\cal A}$
represents nothing but the disorder component which couples to the
dipoles with radius in the shell being integrated out. Since disorder
is uncorrelated on different length scales, we can average over
disorder on a given scale right when we consider screening by dipoles
with a radius of just the same scale. For this reason we can {\em undo
  the replica trick} in the flow equations in the very same way as we
introduced it initially in the full unrenormalized system. In Eq.
(\ref{sum.y}) the replica number $n$ became an explicit variable and
we may now perform the analytic continuation $n\to0$,
\begin{equation} \label{flow.F.n=0}
  \frac{d}{dl} {\cal F} = 2{\cal F} - \pi \ e^{4l} \ \overline{ \ln z}
  .
\end{equation}
As shown in Appendix \ref{app.extreme}, this yields a well defined
free energy per replica only because we have retained {\em all}
classes of replica-vortices. This means, that types of fluctuations,
which seem to be {\em irrelevant} for $n \geq 1$ (in our case:
energetically expensive replica-dipoles) can become {\em important} in
the replica limit.

In a similar way we can proceed to determine the effective fugacities.
In the independent dipole approximation (\ref{Z.lo}) the rescaled
correlation functions read (for $R \geq 1$)
\begin{equation} \label{corr.n}
  \langle N_{\bf R}^a N_{\bf 0}^b \rangle_n = -\sum_\nu y_\nu^2
  N^a_\nu N^b_\nu R^{-2 \pi K_\nu}.
\end{equation}
We ignored normalization by the factor $Z^{(n)}$, since this factor becomes
unity for $n=0$. With the help of a generating function
\begin{equation} \label{def.gen.zeta}
  \zeta(\{\eta \}):= e^{4l} \sum_{\{N^a\}} \overline{ e^{ - 2 E \sum_a
      \left( N^a \right)^2 + \sum_a (2 {\cal A} \sqrt{2 \hat{E}} +
      \eta^a) N^a } }
\end{equation}
we find ($a \neq b$)
\begin{mathletters}\label{undo.corr}
\begin{eqnarray}\label{undo.corr.a}
  \langle N^a_{\bf 1} N^a_{\bf 0} \rangle_n &=& -\left. \frac {d^2
      \zeta(\{\eta\})}{d \eta^a d \eta^a} \right|_{\eta=0} = - e^{4l}
  \ \overline{ \frac{(z_+ + z_-)}{z^{1-n}}} , \\ \langle N^a_{\bf 1}
  N^b_{\bf 0} \rangle_n &=& -\left. \frac {d^2 \zeta(\{\eta\})}{d
      \eta^a d \eta^b} \right|_{\eta=0} = - e^{4l} \ \overline{
    \frac{(z_+ - z_-)^2}{ z^{2-n}}} .
\end{eqnarray}
\end{mathletters}
The $R$-dependence of correlations can be restored by the
substitutions $E \to E+ \pi K \ln R$ and $\hat{E} \to \hat{E}+ \pi
\hat{K} \ln R$. As before, $n$ became an explicit variable such that
we can send $n \to 0$. The effective fugacities then read
\begin{mathletters}\label{undo.yeff}
\begin{eqnarray}\label{undo.yeff.a}
  y^2&=& \frac 12 e^{4l} \ \overline{\frac{z_+ + z_-}{1+z_+ + z_-}} ,
  \\ y_{\rm dis}^2 &=& \frac 12 e^{4l} \ \overline{\left(\frac{z_+ -
        z_-}{1+z_+ + z_-} \right)^2} , \\ y_{\rm con}^2 &=& \frac 12
  e^{4l} \ \overline{\frac{z_+ + z_- + 4 z_+ z_-}{(1+z_+ + z_-)^2} } .
\end{eqnarray}
\end{mathletters}

Now we have achieved our goal of expressing the effective fugacities
of the disordered system ($n=0$) as functions of $E$ and $\hat{E}$. In
principle these fugacities could also depend on $K$ and $\sigma$.
Such a dependence is absent in the independent dipole approximation
used above: correlations are determined neglecting interactions
between dipoles, and the energy of single dipoles of radius unity does
not depend on $K$ and $\sigma$.

Equations (\ref{partial.screen.ren}), (\ref{flow.E}), and
(\ref{undo.yeff}) form a closed set of equations. The flow of free
energy (\ref{flow.F.n=0}) does not feed back into the other flow
equations.  Appealing to the two-component picture, we may call $y^2$
``density of dipoles'', $f:= e^{4l} \overline{ \ln z} / 2 y^2$
``reduced free energy per dipole'', $p:=y_{\rm con}^2/y^2$ ``fraction
of polarizable dipoles'', and $q:=1-p=y_{\rm dis}^2/y^2$ ``fraction of
frozen dipoles''. These quantities are given by Eq. (\ref{undo.yeff})
as functions of $E$ and $\hat{E}$. For the convenience of the reader
we summarize the flow equations with the two-component parameters:
\begin{mathletters}\label{flow.compact}
\begin{eqnarray}\label{flow.compact.F}
  \frac{d}{dl} {\cal F} &=& 2{\cal F}- 2 \pi f y^2 , \\ \frac{d}{dl}
  K^{-1} &=& 4 \pi^3 p y^2 ,
\label{flow.compact.K}\\
\frac{d}{dl} \sigma &=& 4 \pi^3 q y^2 , 
\label{flow.compact.sigma}\\
\frac{d}{dl} E &=& \pi K , \\ \frac{d}{dl} \hat{E} &=& \pi \sigma K^2
.
\end{eqnarray}
\end{mathletters}
As initial values for $K$ and $\sigma$ one has to use the
unrenormalized values, which also enter $E=\pi \gamma K$, $\hat{E}=
\pi \gamma \sigma K^2$ and thereby $y$, $f$, $p$, and $q$.  Initially
${\cal F}(l=0)=0$ since no fluctuations are included.

\subsection{Asymptotic approximation}

In the present form the flow equations are not very convenient since
the dependence of the effective fugacities and the two-component
parameters on $E$ and $\hat{E}$ is still quite intricate. Therefore we
perform an additional approximation which is valid on large length
scales.

First the integrals over ${\cal A}$ (which are to be performed as
disorder averages in Eqs. (\ref{undo.yeff})) are split into intervals
where one of the contributions to the shell partition sum dominates.
Then the normalizing denominators in Eqs. (\ref{undo.yeff}) can be
expanded with respect to the smaller contributions. The resulting
infinite series can be integrated over ${\cal A}$ analytically term by
term. From these series one then can extract easily the leading terms
for large $l$ after approximating $E \approx \pi K l$ and $\hat{E}
\approx \pi \sigma K^2 l$. Since $\frac{d}{dl} E= \pi K$ and
$\frac{d}{dl} \hat{E}= \pi \sigma K^2$, these approximations are
asymptotically correct provided $K$ and $\sigma$ converge to finite
values.

For convenience we introduce an effective temperature variable
\begin{equation} \label{def.tau}
  \tau:= \frac{1}{2 \sigma K}.
\end{equation}
The asymptotic approximation yields for the flow of the reduced free
energy
\begin{equation} \label{flow.F.asy}
\frac{d{\cal F}}{dl}  
 \approx \left\{
\begin{array}{ll}
  2 {\cal F} - 2 \pi e^{ -2 \pi K(1-1/2 \tau) l+4l} & (\tau >1) \\ 2
  {\cal F} - \sqrt{\frac{2 \sigma}{l}} \frac{\pi}{\sin(\pi \tau)}
  e^{-(\pi/2 \sigma)l + 4l} & (\tau <1)
\end{array}
\right.
.
\end{equation}
For the fugacities we find analogously
\begin{mathletters}\label{y.asy}
\begin{eqnarray}\label{y.asy.a}
  y^2 & \approx& \left\{
\begin{array}{ll}
e^{ -2 \pi K(1-1/2 \tau) l+4l}   & (\tau >1) \\
\sqrt{\frac{\sigma}{2 \pi^2 l}} \frac {\pi \tau}{\sin(\pi \tau)} 
e^{-(\pi/2 \sigma)l  + 4l}   & (\tau <1) 
\end{array}
\right. 
, \\
y_{\rm dis}^2 
& \approx& \left\{
\begin{array}{ll}
e^{-4 \pi K l+4l} \left(e^{8 \pi \sigma K^2}-1 \right)  \quad & (\tau >2)  \\
\sqrt{\frac{\sigma}{2 \pi^2 l}} \frac {\pi \tau (1 - \tau)}{\sin(\pi \tau)} 
e^{-(\pi/2 \sigma)l  + 4l}   & (\tau <2) 
\end{array}
\right. 
, \\
y_{\rm con}^2 
& \approx& \left\{
\begin{array}{ll}
e^{ -2 \pi K(1-1/2 \tau) l+4l}   & (\tau >1) \\
\sqrt{\frac{\sigma}{2 \pi^2 l}} \frac {\pi \tau^2}{\sin(\pi \tau)}
e^{-(\pi/2 \sigma)l  + 4l}   & (\tau <1) 
\end{array}
\right. 
.
\end{eqnarray}
\end{mathletters}

Although we were starting from {\em unique} expressions, the {\em
  asymptotic} behavior differs for high temperature or weak disorder
and for low temperature or strong disorder. There are {\em three
  regimes} of parameters, namely $\tau>2$ (called regime IA), $2> \tau
>1$ (called regime IB), and $\tau <1$ (called regime II). Their
physical properties will be analyzed in the next section. 

Approaching the separatrices $\tau=1$ or $\tau=2$ using the
expressions for the high- and low-temperature regime, the expressions
should coincide. This is true for their exponents. The comparison of
the prefactors is not meaningful, since in the low-temperature
expression the diverging factor $1/\sin(\pi \tau)$ is
multiplied with the asymptotically vanishing factor $1/\sqrt{l}$.

Remarkably, for zero temperature ($\tau=0$) we find
$y^2 = y_{\rm con}^2 = R^4 P $, which means that the probability $P$,
Eq. (\ref{prob.dipole}), coincides with the ordinary and connected
correlation.

For the analysis of the flow equations, we wish to use the quantities
of the two-component picture, as introduced in Eqs.
(\ref{flow.compact}).  The reduced free energy per dipole, fraction of
polarizable dipoles, and fraction of frozen dipoles converge
asymptotically to
\begin{mathletters}\label{def.fpq}
\begin{eqnarray}\label{def.f}
f&:=& e^{4l} \ \frac{\overline{\ln z}}{2 y^2} \approx  \left\{
\begin{array}{ll}
1 \quad &  (\tau >1)\\
{1}/{\tau}\quad & (\tau <1)
\end{array}
\right.
, \\
p&:=& \frac{y_{\rm con}^2}{y^2} \approx  \left\{
\begin{array}{ll}
1 \quad & (\tau >1)\\
\tau \quad & (\tau <1)
\end{array}
\right.
\label{def.p}
, \\
q&:=& \frac{y_{\rm dis}^2}{y^2} \approx  \left\{
\begin{array}{ll}
0 \quad & (\tau >1) \\
1-\tau \quad & (\tau <1)
\end{array}
\right.
.
\label{def.q}
\end{eqnarray}
\end{mathletters}
Here we have neglected contributions which vanish on large scales. For
example $q$ is not strictly zero but vanishes exponentially with $l\to
\infty$ for $\sigma >0$ and $\tau>1$. Remarkably, the asymptotic
two-component parameters depend on $\hat{K}$ and $\sigma$ only through
the effective temperature $\tau$.

In order to eliminate $E$ and $\hat{E}$ completely from the asymptotic
flow equations, we determine the flow equation for the fugacity:
\begin{eqnarray}\label{flow.y.asy}
  \frac{d}{dl} y^2 &=& 2y + \pi K \frac{ \partial y}{\partial E} + \pi
  \hat{K} \frac{ \partial y}{\partial \hat{E}} \nonumber \\ &\approx&
  \left\{
\begin{array}{ll}
(4-2\pi K (1- \sigma K)) y^2 \quad & (\tau >1)\\
\left(4 - \frac{\pi}{2\sigma} \right) y^2 \quad & (\tau<1) 
\end{array}
\right. 
.
\end{eqnarray}

To close this subsection of the paper, we wish to point out, that the
flow equations (\ref{flow.compact}) for $K$ and (\ref{flow.y.asy}) for
$y$ with expression (\ref{def.p}) for the polarizability coincide with
those of Nattermann {\em et al}\cite{NSKL95} apart from factors
smaller than 2. This small deviation is due to a rougher asymptotic
approximation in Ref. \CITE{NSKL95}.

\section{Results}

In the previous sections the technical part of deriving the flow
equations has been achieved. The most important flow equations are
Eqs. (\ref{flow.compact}a-c). Originally, the two-component parameters
$f$, $p$, and $q$ have been functions of $E$ and $\hat{E}$. In the
asymptotic approximation $E$ and $\hat{E}$ have been eliminated, the
two-component parameters are given by Eq. (\ref{def.fpq}) as simple
functions of $\tau=1/2\sigma K$, and the fugacity flows according to
Eq. (\ref{flow.y.asy}). These flow equations can be summarized by
\begin{mathletters}\label{flow.sum}
\begin{eqnarray}\label{flow.sum.a}
\frac{d}{dl} y^2 &=& (4-2 \pi \tau^* K (1- \sigma \tau^* K)) y^2 \\
\frac{d}{dl} K^{-1} &=& 4 \pi^3 \tau^* y^2 \\
\frac{d}{dl} \sigma &=& 4 \pi^3 (1- \tau^*) y^2 \\
\frac{d}{dl} {\cal F} &=& 2 {\cal F} - 2 \pi \frac{1}{\tau^*} y^2\\
\tau^* &:=& \min (1,\tau)
\end{eqnarray}
\end{mathletters}
Thus we are in position to discuss the physical implications of these
flow equations.

\subsection{Qualitative aspects}

To start with, we address qualitative aspects and show that the flow
equations satisfy some fundamental physical criteria.

According to Eqs. (\ref{def.y.eff}) and (\ref{undo.yeff}) all
correlations $C_{X}$ are negative, since all fugacities are positive
$y_{X}^2 \geq 0$. This is consistent with a screening that weakens the
attractive interaction between vortices of the same sign and weakens
the repulsive interaction between vortices of the same sign. 

As a consequence, the two-component parameters are also non-negative.
Our asymptotic expressions show, that the fraction of polarizable
dipoles $p=\tau^*$ vanishes linearly at small temperature, whereas the
fraction of frozen dipoles $q=1-\tau^*$ becomes unity.

Screening reduces the coupling $J=KT$ since $\frac {d}{dl} K \leq 0$,
and it increases the strength of disorder, $\frac{d}{dl} \sigma \geq
0$. In comparison to the pure case the screening of $J$ is reduced by
the factor $0\leq p \leq 1$, the fraction of polarizable dipoles.
Whereas previous theoretical approaches did not yield a
renormalization of $\sigma$, Eq. (\ref{flow.sum}c) manifests
that our renormalization scheme leads to an {\em increase} of
$\sigma$. The free energy density is reduced by screening, see Eq.
(\ref{flow.sum}d). Due to disorder the contribution of dipoles is
modified by a factor $f=1/\tau^* \geq 1$.

The increase of $\sigma$ seems to contradict the tendency of thermal
vortices $N$ to {\em neutralize} quenched vortices $Q$. This
neutralization is expected from Eq. (\ref{H.eff}). However, the
screened $\sigma$ is {\em not} defined by the fluctuations of $Q-
\langle N \rangle$. Rather, it was introduced via the fluctuation of
the screened disorder potential $V$. The fluctuations of $V$ are
proportional to $\sigma K^2$. The expected neutralization shows up as
$\frac {d}{dl} \sigma K^2 \leq 0$. This holds even though
$\frac{d}{dl} \sigma \geq 0$ since the reduction of $K$ dominates the
increase of $\sigma$, which follows from $ 2 \sigma K p \geq q$ that
is easily proven with the asymptotic expressions.

The most important quantity to analyze is the flow of entropy.  From
spin-glass models one knows, that a replica-symmetric theory might
lead to a negative entropy. This unphysical feature then indicates
that the symmetric approach is insufficient and one has to allow for a
breaking of replica symmetry. We now show that the present
replica-symmetric theory does not suffer from this problem. Therefore
we believe that a {\em breaking of replica symmetry is not necessary}.

When the entropy ${\cal S}=-\frac{d}{d T} T {\cal F}$ is derived from
the reduced free energy, the unrenormalized values of $J$ and $\sigma$
have to be kept fixed. According to the flow equation for ${\cal F}$,
a non-negative entropy is guaranteed by $\frac{d}{dT} \ln(T f y^2)
\geq 0$. The demonstration of this inequality is complicated by the
temperature dependence, which is implicit to the screened $J$ and
$\sigma$ for $l>0$. If one ignores this implicit dependence, one
easily verifies $\frac{\partial}{\partial T} \ln (T f y^2) \geq 0$,
since $\frac{\partial}{\partial T} T f \geq 0$ and
$\frac{\partial}{\partial T} y^2 \geq 0$ hold for $\tau>1$ and
$\tau<1$ separately. The implicit temperature dependencies can be
taken into account by showing $\frac{\partial}{\partial T} \frac
{d}{dl} \ln (T f y^2) \geq 0$. This is verified in the leading order,
where one has $\frac {d}{dl} \ln (T f y^2)=(4-2 \pi K(1-\sigma K)) +
{\cal O}(y^2)$ for $\tau >1$. If one would use the same expression for
$\tau<1$, one would obtain a negative entropy flow. However, for
$\tau<1$ we find $\frac {d}{dl} \ln (T f y^2)=(4-\frac{\pi}{2
  \sigma}) + {\cal O}(y^2)$. Therefore the entropy flow is
non-negative even at very low temperatures.

\subsection{Ordered phase}

A numerical integration of the flow equations (even without asymptotic
approximations) shows, that below some transition line $L_{\rm c}$ in
the $(T/J, \sigma)$-plane the renormalization flow indeed does
converge for $l \to \infty$ to fixed points with $K_{\infty}$,
$\sigma_{\infty}$, and $y_X=0$. This is true not only in the limit of
large $\gamma$ (small fugacity), but also for arbitrarily small
$\gamma$. Only the extent of the ordered phase shrinks with increasing
$\gamma$. This implies in particular, that the original XY-model
described by $\gamma \approx 1.6$ has an ordered phase. Above $L_{\rm
  c}$ all quantities diverge under renormalization and the present
renormalization group leaves its range of validity, which is limited
to small fugacities already due to the independent dipole
approximation.

The properties of the ordered phase are evident in terms of the
renormalized parameters $K_\infty$ and $\sigma_\infty$. Due to
screening they are related to the bare parameters by $K_\infty \leq K$
and $\sigma_\infty \geq \sigma$, where the inequalities become
equalities for vanishing fugacity within the ordered phase.

\begin{figure}
\epsfxsize=\linewidth
\epsfbox{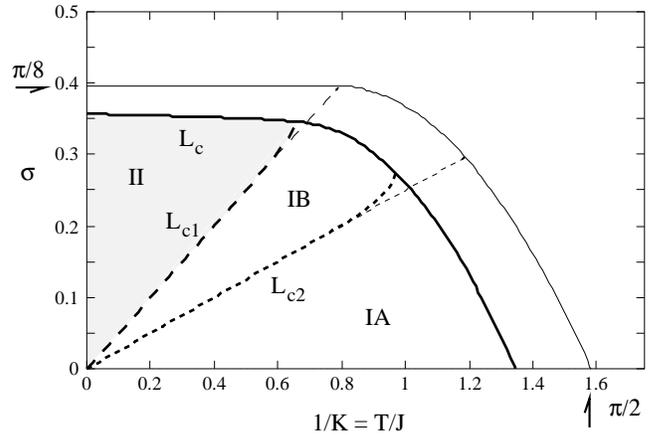}
\caption{
  Phase diagram for $\gamma =1.6$ (thick lines) and $\gamma =10$ (thin
  lines). The transition $L_{\rm c}$ (continuous lines) separates the
  low-temperature phase with quasi long-range order from the
  disordered high temperature phase. The crossover line $L_{\rm c1}$
  (long dashed lines) separates the glassy region II from the
  non-glassy region I. On the crossover line $L_{\rm c2}$ (short
  dashed lines) separates regimes IA and IB within phase I. There the
  disconnected vortex density correlation function changes its
  asymptotic behavior.}
\label{fig_pdg0}
\end{figure}

In the ordered phase, the irrelevance of vortices requires that the
fugacities vanish on large scales. Therefore the flow equation
(\ref{flow.sum}a) of fugacity must have a negative eigenvalue. This is
true at temperatures and disorder strength below $L_{\rm c}$, which is
located at
\begin{equation} \label{sc}
\sigma_\infty= \left\{ 
\begin{array}{ll}
  \frac 1{K_\infty} \left(1-\frac{2}{\pi K_\infty} \right) \quad & (2/
  \pi \leq K_\infty \leq 4/ \pi)\\ {\pi}/{8} \quad &(K_\infty \geq 4/
  \pi)
\end{array}
\right. 
\end{equation}
in terms of the renormalized parameters. It is depicted in Fig.
\ref{fig_pdg0} as function of the renormalized parameters and of the
unrenormalized parameters for $\gamma=1.6$.

The most striking feature of the phase boundary is the absence of
reentrance. Reentrance was predicted by earlier
approaches.\cite{RSN83,CO82} Absence of reentrance without a loss of
the ordered phase was found only recently in improved
approaches.\cite{NSKL95,CF95}

As a function of the renormalized parameters, the phase boundary
$\sigma_\infty (K_\infty)=\pi/8$ has a horizontal part at large
$K_\infty$, i.e. at low temperatures. In terms of bare parameters, the
line $L_{\rm c}$ can be parametrized by a function $\sigma_{\rm c}(K)$
or $K_{\rm c}(\sigma)$. Due to the renormalization of disorder
strength, $\frac{d}{dl} \sigma > 0$ and due to the reduction of the
coupling $\frac{d}{dl} K < 0$ the end points are found at $\sigma_{\rm
  c}(T=0) < \pi/8$ and $K_{\rm c}(\sigma=0) > 2/ \pi$ and $\sigma_{\rm
  c}(T)$ has no longer a horizontal part for finite fugacity. 

Concerning the asymptotic behavior of the vortex correlation
functions, one can distinguish three regions within the ordered phase.
They are separated by a line $L_{\rm c1}$, fixed by $2 \sigma_\infty
K_\infty =1$ (equivalent to $\tau_\infty=1$), and a line $L_{\rm c2}$,
fixed by $4 \sigma_\infty K_\infty =1$ (equivalent to
$\tau_\infty=2$), see Fig. \ref{fig_pdg0}. In the
$(J/T,\sigma)$-diagram these lines emerge from the origin. They are
straight lines in the limit of zero fugacity and bent upwards for
finite fugacity.

The correlations are retrieved from the fugacities,
Eq. (\ref{y.asy}), by omitting rescaling and inserting $l= \ln R$.
Thereby we find ($R \gg 1$)
\begin{mathletters}\label{Cs}
\begin{eqnarray}\label{C_con}
  C({\bf R}) &\approx& \left\{
\begin{array}{l}
  \displaystyle -2 R^{- 2\pi K_{\infty}(1-1/2\tau_{\infty})} \\ 
  \displaystyle -2 \sqrt{\frac {\sigma_{\infty}}{2 \pi^2 \ln R}} \ 
  \frac{\pi \tau_\infty}{\sin( \pi\tau_{\infty})} \ R^{- \pi/2
    \sigma_{\infty}}
\end{array}
\right.  \\ C_{\rm con}({\bf R}) &\approx& \left\{
\begin{array}{l}
  \displaystyle -2 R^{- 2\pi K_{\infty}(1-1/2\tau_{\infty})} \\ 
  \displaystyle -2 \sqrt{\frac {\sigma_{\infty}}{2 \pi^2 \ln R}} \ 
  \frac{\pi \tau_\infty^2}{\sin( \pi\tau_{\infty})} \ R^{- \pi/2
    \sigma_{\infty}}
\end{array}
\right.
\end{eqnarray}
for $\tau_\infty {\mbox{\hbox{ \lower-.6ex\hbox{$>$}\kern-.8em
      \lower.5ex\hbox{$<$}\kern+.35em}}} 1$ and
\begin{equation} \label{C_dis}
  C_{\rm dis}({\bf R}) \approx \left\{
\begin{array}{l}
  \displaystyle -2 R^{- 4\pi K_{\infty}} \left( R^{4 \pi
      K_{\infty}/\tau_{\infty}}-1 \right) \\ \displaystyle -2
  \sqrt{\frac {\sigma_{\infty}}{2 \pi^2 \ln R}} \ \frac{\pi
    \tau_\infty (1-\tau_\infty )}{\sin( \pi\tau_{\infty})} \ R^{-
    \pi/2 \sigma_{\infty}}
\end{array}
\right.
\end{equation}
\end{mathletters}
for $\tau_\infty {\mbox{\hbox{ \lower-.6ex\hbox{$>$}\kern-.8em
      \lower.5ex\hbox{$<$}\kern+.35em}}} 2$. The limiting values for
the fraction of polarizable and frozen dipoles are related to the
correlations by
\begin{mathletters}\label{pqf.inf}
\begin{eqnarray}\label{p.inf} 
  p_{\infty}&=&\lim_{R\to \infty} \frac{C_{\rm con}({\bf R})}{C({\bf
      R})} , \\ q_{\infty}&=&\lim_{R\to \infty} \frac{C_{\rm dis}({\bf
      R})}{C({\bf R})} . 
\end{eqnarray}
\end{mathletters}

Naturally the question arises, whether these regimes are separated by
a true thermodynamic transition at $L_{\rm c1}$ and $L_{\rm c2}$. As
we have observed, that asymptotic form of the flow equations and of
the correlations changes its dependence on $\sigma$ and $K$
non-analytically at $\tau=1$ and $\tau=2$. However, a phase transition
in the strict sense requires a non-analytic dependence of the
integrated free energy on the unrenormalized parameters $\sigma$ and
$K$. We were not able to prove such a singularity at $L_{\rm c1}$ and
$L_{\rm c2}$, which we therefore consider as crossover lines only.
However, a true transition is expected at $L_{\rm c}$ just as
in the pure case.

Region II can be distinguished {\em qualitatively} from the other
regions.  One can consider $q_{\infty}$ as order parameter for the
low-temperature regime II. It measures the relative strength of the
disconnected vortex density correlation to the usual correlation on
large scales. In the definition of this order parameter the
normalization of the disconnected correlation by the usual correlation
is necessary, since true long-range order is absent.  The glassy order
reflected by the Edwards-Anderson like correlation can therefore be
only quasi long-ranged, too. Nevertheless one may compare $q_{\infty}$
to the Edwards-Anderson order parameter in spin-glasses and we
interpret it as order parameter for the {\em glassiness} of in the
vortex system.

The regions IA and IB differ only by the analytic dependence of the
connected correlation on $K_\infty$ and $\sigma_\infty$. The
correlation under consideration do not provide an order parameter
comparabe to $q$ for the line $L_{\rm c2}$.

\subsection{Universality}

The universality class of the pure XY-model is defined by the behavior
of the critical exponents $\eta=1/4$ and $\delta=15$ at the transition
and the exponential divergence of the correlation length when the
transition is approached from above.\cite{jmK74}

In the presence of disorder, the spin correlation decays with
exponent\cite{RSN83}
\begin{equation} \label{eta.ren}
  \eta_{\infty}=(1/K_\infty+\sigma_\infty) /2\pi.
\end{equation}
Along the line $L_{\rm c}$ the exponent takes the value $\eta_\infty =
1/4$ for $\sigma=0$,\cite{jmK74} which decreases with temperature
until it reaches the value $\eta_\infty = 1 /16$ for
$T=0$.\cite{NSKL95} In the disordered phase, where correlations decay
exponentially, $\eta_\infty =\infty$ formally. Thus the value of
$\eta_\infty$ at the transition becomes non-universal, i.e. it is
disorder dependent. Only for $\sigma =0$ or $T=0$ the jump of
$\eta_\infty$ is universal (from 1/4 or 1/16 to $\infty$ respectively)
in the sense that it does not depend on the bare coupling $J$ or bare
fugacity $y$.  For $\sigma >0$ or $T>0$ the jump lies inbetween and
will depend on microscopic details.

In the present work we have not included a magnetic field. Therefore
we can not determine $\delta$ in the presence of disorder.

We now attempt to analyze how the correlation length $\xi_+$ diverges
when the transition line is approached from above. Even though the
flow equations become invalid for large fugacity,
Kosterlitz\cite{jmK74} has argued that $\xi_+$ is given by the length
scale where the fugacity becomes of order unity. In the pure case,
where the flow equations reduce to $\frac {d}{dl} K^{-1} = 4 \pi^3
y^2$ and $\frac {d}{dl} y^2 = (4- 2 \pi K) y^2$ he found $\xi_+ \sim
\exp ( b/ \sqrt{K^{-1}- K_{\rm c}^{-1}})$ with a non-universal
constant $b$.  In the zero-temperature limit the flow equations read
$\frac {d}{dl} \sigma = 4 \pi^3 y^2$ and $\frac {d}{dl} y^2 = (4-
\frac{\pi}{2 \sigma}) y^2$. By a replacement $\sigma \to 1/4K$ they
can be mapped onto the flow equations of the pure case. Therefore we
conclude, that the zero temperature transition has
\begin{equation} \label{div.xi}
  \xi_+ \sim \exp ( b / \sqrt{\sigma - \sigma_{\rm c}})
\end{equation}
with some other non-universal $b$. For finite temperature {\em and}
finite disorder, our flow equations are too complicated for being
integrated up analytically. However, we naturally expect that the
functional form of the divergence is preserved and that only the
argument of the square-root has to be replaced by the distance to the
transition line, say $\min_{(\sigma_{\rm c},K_{\rm c}) \in L_c}
[(K^{-1}- K_{\rm c}^{-1})^2 + ( \sigma - \sigma_{\rm c})^2]^{1/2}$.

\section{Discussion and conclusions}

The present theory is based on a selfconsistent linear screening
approach, which is solved in a differential form. This differential
form has been presented in the renormalization group language. In
fact, the very same differential equations can be directly derived by
a real space renormalization group, generalizing the scheme of
Kosterlitz\cite{jmK74} to the replicated system. In this scheme the
flow equations obtain contributions from integrating out {\em all}
classes of replica dipoles with the smallest (cut-off) diameter.

The controversy among earlier theoretical treatments of the disordered
XY-model requires to put the present work into perspective with these
earlier approaches.

When Cardy and Ostlund (CO) considered the more complicated problem of
the XY-model in the presence of random fields,\cite{CO82} they used a
replica-symmetric approach to derive renormalization group flow
equations, which generated also random bond disorder. Rubinstein,
Shraiman, and Nelson\cite{RSN83} (RSN) focused on the random bond
problem, which they analyzed without using the replica formalism.
However, the resulting flow equations were a subset of those of CO.
Their starting point was essentially Eq. (\ref{partial.screen.K}),
which they evaluated only to the leading order in the fugacity of
vortices in the unreplicated system. Consequently, they neglected the
disconnected correlation and could not find a flow of $\sigma$. Their
flow equations for $K$ and $y^2$ coincide with our asymptotic flow
equations for $\tau >1$. However, they used them also for $\tau <1$,
which gave rise to reentrance and a negative flow of entropy.

The present approach is in principle equivalent to CO and RSN. The
essential difference lies in the calculation of the correlations, or
in the contributions which are taken into account when the dipoles of
smallest distance are integrated out. CO and RSN thereby made too
crude approximations, which treat the disorder effectively as
annealed. This yields correct results only for high temperatures $\tau
>1$ but fails for $\tau <1$.

Technically, they considered only contributions of replica-vortices of
class $(m_0=1,m_1=0,1)$. We pointed out, that already for replica
numbers $n >1$ this restriction is insufficient at low temperatures.
Therefore we included {\em all classes} of replica-vortices. The
additional vortex classes turned out to be important for $\tau <1$ and
remove reentrance and the negative flow of entropy.

From the disappearence of reentrance after the inclusion of additional
vortex classes we have to conclude, that these {\em additional}
fluctuations in the replicated system ($n>1$) {\em reduce} the effect
of thermal fluctuations in the original unreplicated system ($n=0$).
This strange feature reminds of the fact, that in variational replica
approaches one has to maximize and not to minimize the free energy for
$n=0$ in order to find the physical state.\cite{minmax}

When Korshunov\cite{seK93} considered the problem in the replica
representation, he attempted to identify the transition in the limit
of zero fugacity with the dissociation of the least stable class of
dipoles. The $N$-complexes in his notation are identical to dipoles
formed by replica vortices of class $(m_0,m_1=0,m_0)$, where one has
to identify $N \equiv m_0$. They indeed drive the transition for $n
\geq 1$.  From my point of view, there are problems in his way of
taking the limit $n \to 0$, which cause the conclusion, that the
ordered phase should not exist at all. One problem is, that it does
not make sense to consider replica vortices which occupies $m_0>1$
replicas even for a number of replicas $n<1$, since $(n-m_0)(m_0-1)
\geq 0$ should be preserved during the analytic continuation.

In addition, the consideration of such restricted classes of dipoles
is not sufficient for an analytic continuation $n \to 0$. Even then,
if one takes into account all classes of replica-vortices, it is the
{\em most stable} and not the least stable class, which drives the
transition in the limit $n \to 0$. This is again a strange feature of
replica theory. The arguments for these statements are detailed in
Appendix \ref{app.extreme}.

In Appendix \ref{app.sin.vor} we further show, that $L_{\rm c}$ can be
derived already from considering a {\em single} vortex in a finite
system. After replication its free energy can be calculated by taking
into account certain types of fluctuations, which are again
parametrized by some $m$ in the range $(n-m)(m-1) \geq 0$. The
transition $L_{\rm c}$ follows from a {\em maximization} of the free
energy with respect to this parameter.

For several disordered systems the occurence of reentrance and
negative contributions to the flow of entropy have been shown to be
artefacts of a replica-symmetric theoretical approach, which could be
removed by breaking replica symmetry.\cite{RSB+reen} For the present
model these artefacts have emerged in the early replica-symmetric
approaches, indeed. Here we have shown that these problems can be
cured by including the relevant fluctuations within a symmetric
theory. Hence a breaking of replica symmetry is not necessary.

The recent approaches of Nattermann, Scheidl, Korshunov, and
Li\cite{NSKL95,KN95} and also of Cha and Fertig\cite{CF95} found the
existence of the ordered phase and the absence of reentrance without
using the replica formalism. Their results agree with the present
theory at least in the limit of zero fugacity, for finite fugacity
there are quantitative deviations. In a very recent real-space
renormalization group approach in the unreplicated system,
Tang\cite{lhT95} has derived flow equations identical to Eqs.
(\ref{flow.sum}) and achieves even quantitative agreement with the
present results.

A qualitative progress of the present theory is the inclusion of a
renormalization of $\sigma$. This renormalization is driven by the
disconnected vortex density correlation, which in earlier approaches
was neglected from the very beginning or found to be zero due to later
approximations. Although $\sigma$ is increased under renormalization,
this effect does not destroy the ordered phase. The flow of $\sigma$
is essential for the universality type of the zero-temperature
transition, which deviates from the Kosterlitz-Thouless transition in
the pure system only by the value of the critical exponents but not in
the exponential divergence of the correlation length.

Due to the renormlaization of $\sigma$ one can exclude a horizontal
part of $L_{\rm c}$ at low temperatures even for finite fugacity. Such
a horizontal part was shown to exist by Ozeki and Nishimori\cite{ON93}
for a wide class of disordered systems with a certain gauge symmetry.
But the XY-model under consideration does not possess this symmetry.

To sum up, is has been shown that the two-dimensional XY-model with
random phase-shifts on bonds has a phase with quasi long-range order.
It has exhibits a zero temperature transition at a critical strength
of disorder. Near the transition the correlation length diverges
exponentially as in the pure case, but the jump of the spin
correlation exponent is of different magnitude.

Within the ordered phase there is a glassy low-temperature region. In
the absence of true long-range order, we identified the ratio of the
Edwards-Anderson correlation to the usual vortex-density correlation
on large length scales as order parameter for glassiness.

We used a replica-symmetric renormalization group approach. The
reentrance and negative entropy flow of earlier approaches of that
kind could be cured by taking into account additional fluctuations. No
need for a breaking of replica symmetry was found.

Several generalizations of the present work are of interest. In a
forthcoming publication,\cite{prep} the present theory will be
extended to include (random) magnetic fields on the ordered phase of
the XY-model. The properties of the system in the disordered phase are
beyond the range of our perturbative renormalization approach. In
particular the question about possible glassy properties in this phase
deserves further research. In the limit $\sigma \to \infty$, the
present model becomes the so-called gauge glass model, which also
exhibits a glass transition.\cite{gauge_glass}

\section*{Acknowledgments}

The author gratefully acknowledges stimulating discussions with S.E.
Korshunov, T. Nattermann, and L.-H. Tang. This work was supported
financially by the Deutsche Forschungsgemeinschaft SFB 341. The author
also enjoyed hospitality of I.S.I. Foundation funded by EU contract
ERBCHRX-CT920020.

\appendix

\section{Linear response}
\label{app.lr}

Since Eqs. (\ref{partial.screen}) are the basis of the present
analysis, it seems worthwhile to present some intermediate steps of
their derivation. We start by including a test vorticity $q$
into the effective vortex Hamiltonian (\ref{Hv.k}),
\begin{equation} \label{app.H_c}
  {\cal H}_{\rm v}= \frac 12 \int_{\bf k} U_{\bf k} \left|N_{\bf
      k}+q_{\bf k}-{Q}_{\bf k} \right|^2 .
\end{equation}
The free energy of the system with test vortices reads
\begin{equation} \label{app.F(nu)}
  {\cal F}[\{ q \}] = -\ln \sum_{\{ N\}} e^{-{\cal H}_{\rm v}}.
\end{equation}
The screened potential and interaction are defined by
\begin{mathletters}\label{app.UV.scr}
\begin{eqnarray}\label{app.U.scr}
V_{\bf k}^{\rm scr} &:=& \left. \frac{ \delta {\cal F} [\{ q \}]}
{\delta q_{-\bf k}} \right|_{q=0} ,  \\
U_{\bf k}^{\rm scr} \delta({\bf k}+{\bf k}')&:=& 
\left. \overline{\frac{ \delta^2 {\cal F} [\{ q \}]}
{\delta q_{-\bf k}\delta q_{-\bf k'}}} \right|_{q=0} ,   
\end{eqnarray}
\end{mathletters}
and result in Eqs. (\ref{screen.V}). The screened parameters are
identified via

\begin{mathletters}\label{app.Ks.scr}
\begin{eqnarray}\label{app.K.scr}
K^{\rm scr} &:=& \lim_{k \to 0} \frac{k^2}{4 \pi^2} \ U_{\bf k}^{\rm scr} ,
 \\
\sigma^{\rm scr}(K^{\rm scr})^2 &:=& \lim_{k \to 0} 
\frac{k^2}{4 \pi^2}  \overline{V_{-\bf k}^{\rm scr} \ V_{\bf k}^{\rm scr}} .
\end{eqnarray}
\end{mathletters}

Fourier transformation then leads to Eqs. (\ref{total.screen}). On the
way one needs the probability distribution of quenched vortices,

\begin{equation} \label{app.P(Q)}
P[\{ Q\}] \propto \exp \left\{- \frac { 2 \pi^2}{\sigma} 
\int_{\bf k} \frac{1}{k^2} |Q_{\bf k}|^2 \right\},
\end{equation}
and one has to integrate partially

\begin{eqnarray}\label{app.QN}
\overline {Q_{\bf k} \langle N_{-\bf k} \rangle}& =& 
\int {\cal D}\{Q\}  P[\{ Q\}] Q_{\bf k} \langle N_{-\bf k}  \rangle 
\nonumber \\
&=&- \frac {\sigma k^2}{4 \pi^2} \int {\cal D}\{Q\}  
\frac{\delta P[\{ Q\}]}{\delta Q_{-\bf k}} \langle N_{-\bf k} \rangle 
\nonumber \\
&=&  \frac {\sigma k^2}{4 \pi^2} \int {\cal D}\{Q\}  P[\{ Q\} ]
\frac{\delta \langle N_{-\bf k} \rangle}{\delta Q_{-\bf k}} \nonumber \\
&=& \sigma K C_{\rm con}({\bf k}) . 
\end{eqnarray}

The resulting screening formulae are also valid in the case, where the
interaction and disorder correlation on small scales have additional
contributions due to $E$ and $\hat{E}$, which flow under renormalization
independently of $K$ and $\sigma$.

It is important to realize, that $\sigma^{\rm scr}$ could
alternatively be defined by $\frac {4 \pi^2}{k^2} \overline{(Q_{-{\bf
      k}}- \langle N_{-{\bf k} }\rangle )(Q_{{\bf k}}- \langle N_{{\bf
      k} }\rangle )}$ for small $k$, which also might be plausible at
first sight but gives different results as soon as $K^{\rm scr} \neq
K$. Then, however, it would be necessary to distinguish the screened
interactions among thermal vortices $N$ from the screened interactions
between thermal vortices $N$ and quenched vortices $Q$. This
complication is avoided by the definition actually used, which is
based on the correlation of $V^{\rm scr}$. Only in this way screening
effects can be captured by a renormalization of the parameters $K$ and
$\sigma$, which were present already in the unrenormalized model.

\section{Extremal principle}
\label{app.extreme}

In this appendix we address the question, whether the correct vortex
flow equations can be found, if one considers only the contributions
of a {\em single} class of replica vortices to the partition sum and
the correlations.

For $n>1$ one naturally expects, that fluctuations of the vortices
with the lowest energy cost give the most important contribution to
renormalization. Since the interaction between different replicas
favors vortices of the same sign, optimal replica vortices will be of
class $(m_0,m_1=0)$ or $(m_0,m_1=m_0)$, which we call simply $m \equiv
m_0$ vortices. According to Eq. (\ref{def.E_m}), their core energy is
\begin{equation} \label{def.E_m.app}
  E_m=mE-m^2 \hat{E} .
\end{equation}
Since $E$ and $\hat{E}$ are positive, the dipoles with minimum energy
are always given by $m=1$ for $(n+1) \hat{E} \leq E$ and by $m=n$ for
$(n+1) \hat{E} \geq E$. They determine the phase diagram of Fig.
\ref{fig_pdgn}. The {\em most stable} pairs are given by
$m^*=E/2\hat{E}$ (or, for $n\geq 1$, by one of the two nearest
integers to $m^*$).

Now we consider screening effects by $m$-vortices only. We would like
to point out, that the restriction to this class of vortices does not
break replica symmetry, since we include all $2 {n \choose m}$
possibilities to place such vortices in $n$ replicas. Then the flow
equations read
\begin{mathletters}\label{total.screen.m}
\begin{eqnarray}\label{total.screen.m.a}
  \frac{d}{dl} {\cal F} &=& 2 {\cal F} - 2 \pi \frac 1n {n\choose m}
  y_m^2 ,\\ \frac{d}{dl} K^{-1}&=& 4 \pi^3 y_{\rm con}^2 , \\ 
  \frac{d}{dl} \sigma &=& 4 \pi^3 y_{\rm dis}^2 ,
\end{eqnarray}
\end{mathletters}
with fugacity $y_m=e^{2l-E_m}$ and effective fugacities $y^2= {n-1
  \choose m-1} y_m^2$, $y_{\rm dis}^2= {n-2 \choose m-2} y_m^2$, and
$y_{\rm con}^2= {n-2 \choose m-1} y_m^2$. They can be related by the
identities ${n \choose m} = \frac {n}{m} {n-1 \choose m-1}$, ${n-2
  \choose m-1} = \frac {n-m}{n-1} {n-1 \choose m-1}$, and ${n-2
  \choose m-2} = \frac {m-1}{n-1} {n-1 \choose m-1}$. Thereby we find
the two-component parameters and flow of $y$
\begin{mathletters}\label{flow.m}
\begin{eqnarray}\label{flow.m.a}
  f &=& \frac 1m , \\ p&=& \frac {n-m}{n-1} , \\ q &=& \frac
  {m-1}{n-1} , \\ \frac{d}{dl} y^2 &=&(4-2 \pi K_m) y^2 ,
\end{eqnarray}
\end{mathletters}
where $K_m:=mK-m^2 \hat{K}$.

To simplify the discussion of the replica-limit $n=0$, we use the
asymptotic approximation $E \approx \pi K l$ and $\hat{E} \approx \pi
\sigma K^2 l$. One should consider $m$ in the range $0\leq m \leq 1$,
which follows from an analytic continuation of the condition
$(n-m)(m-1)\geq 0$.

One easily convinces oneself in the case {\em without} disorder
($\hat{E}=0$), that one obtains unphysical results, if one looks for
$m$ which {\em minimizes} $K_m$: this would be $m=0$ and, the systen
would be unstable to the creation and dissociation of dipoles
($E_0=K_0=0$).  However, if we look for the {\em maximum}, which lies
in the pure case always at $m=1$, we find a positive core energy $E_1$
and the correct phase boundary at $K_1=K=2/\pi$.

Therefore let us try to look for the most stable type of dipoles also
in the presence of disorder. It is given by $m^* =\tau:= 1/2\sigma K$.
Recalling the constraint $0 \leq m^*\leq 1$ in the limit of $n=0$, we
have to distinguish a high- and low-temperature regime with
\begin{mathletters}\label{m^*}
\begin{eqnarray}\label{m^*.a}
  m^*=1; & \quad K_{m^*}= K (1-1/2\tau) & \quad {\rm for} \quad \tau
  \geq 1, \\ m^*=\tau ; & \quad K_{m^*}=1 / 4 \sigma & \quad {\rm for}
  \quad \tau \leq 1.
\end{eqnarray}
\end{mathletters}
If the core energy is large and renormalization is negligible, the
criterion $K_{m^*} \geq 2/ \pi$ for the dissociation of the most
stable dipoles then yields the correct transition line $L_{\rm c}$ as
in Eq. (\ref{sc}). Surprisingly, even the two-component parameters
$f=1/\tau$, $p=\tau$, and $q=1-\tau$ for $\tau <1$ according to Eq.
(\ref{flow.m}) are in perfect agreement with (\ref{def.fpq}).

So, was it essentially unnecessary to consider {\em all} classes of
vortices?  No! Even though the flow equations are correct, we must
know the initial values to integrate them up, in particular we must
know the unrenormalized value of $y$. The problem is, that for $n \to
0$ we have $y^2 = {n-1 \choose m^*-1} y_{m^*}^2 \to \frac{\sin( \pi
  m^*)}{\pi n} y_{m^*}^2$, which {\em diverges} $\sim n^{-1}$ in the
low temperature regime with $0< m^* < 1$. This pretends erroneously
the relevance of vortices. In particular, there is an inconsistency
with the replica trick, since according to Eq. (\ref{total.screen.m}a)
the free energy per replica diverges like $\frac 1n {n \choose
  m^*}=\frac 1{m^*} {n-1 \choose m^*-1} \to \frac{\sin( \pi m^*)}{m^*
  \pi n} \propto n^{-1}$. These problems are removed by including the
contributions of all classes of vortices.

\section{Single Vortex argument}
\label{app.sin.vor}

It is instructive to show, that the phase boundary $L_{\rm c}$ in the
limit of zero fugacity can be derived by considering a single vortex.
Similar to the reasoning of Kosterlitz and Thouless,\cite{KT73} we
examine the possibility of finding single free vortices in a system of
finite linear dimension $L$ by determining its free energy. 

The energy of a vortex is composed of its self-energy $\pi J \ln L$
and of a potential energy $V$ due to disorder. The disorder-averaged
free energy reads
\begin{eqnarray}\label{F.1}
  {\cal F}&=&\pi K \ln L + \delta {\cal F} , \\ \delta {\cal F} &=& -
  \overline{ \ln \int d^2r e^{-\beta V({\bf r})} } = - \lim_{n \to 0}
  \frac 1n \ln \overline{ Z^n } , \\ \overline{ Z^n } &=& \int d^2r_1
  \dots d^2r_n \times \\ & & e^{-\frac 12 \beta^2 \sum_{1 \leq a < b
      \leq n} \Delta^2({\bf r}_a-{\bf r}_b) +\frac n2 \beta^2 \sum_{1
      \leq a \leq n} \overline{V^2({\bf r}_a)} } . \nonumber
\end{eqnarray}
We use $\overline{V^2({\bf r}_a)} \approx 2 \pi \sigma J^2 \ln L$
and find the leading power of $\overline{ Z^n }$ in $L$ by considering
all possible partitionings ${\cal P}$ of replicas into
groups:\cite{HS91}
\begin{equation} \label{Z.n}
  \overline{ Z^n } \sim L^{n^2 \pi \sigma K^2 + \max_{\cal P} \left[2
      |{\cal P}|- 2 \pi \sigma K^2 \sum_{1 \leq a < b \leq n}
      (1-\delta_{a,b}^{\cal P}) \right]} .
\end{equation}
Here $|{\cal P}|$ denotes the number of groups, and
$\delta_{a,b}^{\cal P}=1$ or $0$ if replicas $a$ and $b$ are in the
same or in different groups. If we consider only the possibility,
where $n/m$ groups of $m$ replicas are formed, we find
\begin{equation} \label{Z.n.2}
  \overline{ Z^n } \sim L^{n^2 \pi \sigma K^2 + \max_{m} \left[2 n/m -
      \pi \sigma K^2 n (n-m) \right]} .
\end{equation}
The restriction to these partitionings looks at first sight like
a one-step replica symmetry breaking, but it is not. If one
partitioning is included, also all other partitionings, wich are
related by a permutation of replicas, are also included.

In the analytic continuation $n \to 0$ the group parameter has to be
taken in the range $0 \leq m \leq 1$ and the maximization is replaced
by a minimization. Then
\begin{equation} \label{dF}
  \delta {\cal F} = - \min_{0 \leq m \leq 1} \left[2/m + \pi \sigma
    K^2 m \right] \ln L .
\end{equation}

Finally, for $\pi \sigma K^2<2$ the minimum is taken at $m=1$ and we
obtain
\begin{equation} \label{F.HT}
  {\cal F} = \left\{ \pi K(1- \sigma K)- 2 \right\} \ln L ,
\end{equation}
whereas for $\pi \sigma K^2>2$ the minimum is taken at $m=\sqrt{2/\pi
  \sigma K^2}<1$ and we obtain
\begin{equation} \label{F.LT}
  {\cal F} = \pi K \left\{1- \sqrt{8 \sigma/\pi } \right\} \ln L .
\end{equation}

The phase boundary $L_{\rm c}$ is there, where the free energy of the
single vortex changes sign. Eq. (\ref{F.HT}) gives the
high-temperature part of $L_{\rm c}$, whereas Eq. (\ref{F.LT}) gives
the low-temperature part of $L_{\rm c}$ in agreement with Eq.
(\ref{sc}). Note, that the boundary $\pi \sigma K^2=2$ separating the
validity of Eqs. (\ref{F.HT}) and (\ref{F.LT}) does {\em not} coincide
with $L_{\rm c1}$ of $L_{\rm c2}$. This, however, cannot be expected
since within the ordered phase fluctuations are dominated by vortex
dipoles and not by single vortices. The appearence of the separatrix
$\pi \sigma K^2=2$ is specific for the physics of single vortices.

\end{multicols}

\end{document}